\documentclass[10pt,twocolumn,letterpaper,journal]{IEEEtran}

\usepackage{balance}
\usepackage{cite}
\usepackage{epsfig}
\usepackage{times}
\usepackage{graphicx}
\usepackage[figuresright]{rotating}
\usepackage{subcaption}
\usepackage{amssymb}
\usepackage{amsmath}
\usepackage{setspace}
\usepackage{rotating}
\usepackage{multirow}
\usepackage{wrapfig}
\usepackage{booktabs}
\usepackage{array}
\usepackage{comment}
\usepackage{epstopdf}
\usepackage[utf8]{inputenc}
\usepackage{parskip}
\usepackage{amsmath}
\usepackage{indentfirst}
\usepackage{graphicx}
\graphicspath{ {images/} }
\usepackage[rightcaption]{sidecap}
\usepackage[export]{adjustbox}[2011/08/13]

\usepackage{tabu}
\usepackage{booktabs}%
\usepackage{tabularx}
\usepackage[font=footnotesize]{caption}

\let\OLDthebibliography\thebibliography
\renewcommand\thebibliography[1]{
  \OLDthebibliography{#1}
  \setlength{\parskip}{0pt}
  \setlength{\itemsep}{5pt plus 0.3ex}
}

\makeatletter
\IEEEtriggercmd{\reset@font\normalfont\fontsize{7.9pt}{8.40pt}\selectfont}
\makeatother
\IEEEtriggeratref{1}

\IEEEoverridecommandlockouts

\usepackage{algorithmic}
\usepackage{algorithm}
\floatname{algorithm}{Algorithm} \textfloatsep 0.5cm

\usepackage{colortbl}

\begin{document}
\title{Improved Circuit Design of Analog Joint Source Channel Coding for Low-power and Low-complexity Wireless Sensors}
\author{Xueyuan~Zhao,~\IEEEmembership{Member,~IEEE,}
        Vidyasagar~Sadhu,~\IEEEmembership{Student~Member,~IEEE,}
        Anthony~Yang,~\IEEEmembership{Student~Member,~IEEE,}
        and~Dario~Pompili,~\IEEEmembership{Senior~Member,~IEEE}%
\thanks{The authors are with the Department of Electrical and Computer Engineering, Rutgers University–-New Brunswick, NJ, USA.}
\thanks{Emails:\{xueyuan.zhao,vidyasagar.sadhu,anthony.yang,pompili\}@rutgers.edu}
\thanks{A preliminary version of this work appeared in the \emph{Proc. of the IEEE Intl. Symposium on Circuits \& Systems~(ISCAS)}, Montreal, Canada, May~2016~\cite{Zhao16}.}}%

\maketitle
\thispagestyle{empty}

\begin{abstract}
To enable low-power and low-complexity wireless monitoring, an improved circuit design of Analog Joint Source Channel Coding~(AJSCC) is proposed for wireless sensor nodes. This innovative design is based on Analog Divider Blocks~(ADB) with tunable spacing between AJSCC levels. The ADB controls the switching between two types of Voltage Controlled Voltage Sources~(VCVS). LTSpice simulations were performed to evaluate the performance of the circuit, and the power consumption and circuit complexity of this new ADB-based design were compared with our previous parallel-VCVS design. It is found that this improved circuit design based on ADB outperforms the design based on parallel VCVS for a large number of AJSCC levels ($\geq 16$), both in terms of power consumption as well as circuit complexity, thus enabling persistent and higher temporal/spatial resolution environmental sensing.
\end{abstract}
\begin{keywords}
Power Efficiency, Analog Compression, Circuit Complexity, Signal Multiplexing, Environmental Monitoring.
\end{keywords}

\section{Introduction}\label{sec:introduction}

\textbf{Motivation and Background:}
High temporal/spatial resolution environmental monitoring applications involve deploying a large number of power-efficient and low-complexity/low-cost wireless sensors~\cite{wons3tier2017,Zhao16}. Some application scenarios include low-cost, high-confidence monitoring of urban infrastructure~\cite{Kumar16}, 
intelligent transportation systems~\cite{Hu15}, wireless health monitoring~\cite{iscas17}, 
datacenter monitoring~\cite{Viswanathan11}, ocean environment monitoring~\cite{joe17}.
For such applications, we observe that: i)~sensors need to be deployed in a wide area and with high density, hence a large number of disposable low-cost sensors are needed; ii)~nearby sensors are clustered and transmit wireless signals to a cluster-head node in separate and multiplexed channels; iii)~sensors are expected to be powered by battery or even self-powered
and should work for extended periods of time. All these requirements demand low-power, low-cost/complexity designs of the sensor nodes.

\textbf{State of the Art:}
Current research efforts to reduce the power consumption of wireless sensors include: i)~reducing the power of radio Radio Frequency~(RF) components as well as the power of radio transmission; ii)~adopting new materials to reduce the power needs of sensing components; iii)~adopting energy-harvesting power supplies. To reduce the radio power, a wake-up radio-receiver approach is proposed to save radio power~\cite{Magno16wakeup}. A transmission adaptive power-control approach is proposed based on wireless link-quality measurement for wireless monitoring sensors~\cite{Basu17}. An event-driven approach is described to reduce the power consumption for large-scale wireless monitoring~\cite{Olsson16}. To reduce the power consumption at the sensing components, a new type of carbon nanotube is proposed for low-power gas sensing~\cite{Magno16gas}. Power circuits for energy harvesting wireless sensors are also proposed~\cite{Ahmed16}. These existing research works address reducing power-consumption for digital circuits. The advantage of such digital architecture is that functionalities such as digital data compression, channel coding, and digital modulation (as well as various types of signal processing) can be done in a digital processing chipset. However, the drawback is that Analog-to-Digital-Converter~(ADC), Digital-to-Analog Converter~(DAC), and digital chipsets are needed, all of which are power demanding components, resulting in high circuit complexity and high cost. A digital architecture, therefore, tends to yield powerful yet non-disposable and non-persistent wireless sensing motes.

\textbf{Power Consumption:} 
Digital wireless sensing motes with their microprocessors consume much higher power than our proposed analog circuit. The active microprocessor power consumptions for these sensors (as in~\cite{wons3tier2017}) are: $1.1~\rm{mW}$ for WSN340, $2.4~\rm{mW}$ for Mantaro CoSeN, $6~\rm{mW}$ for Telos RevB, and $26.4~\rm{mW}$ for MICA2. For our proposed ADB design with 64 AJSCC levels, assuming nano fabrication technology of the analog components, the consumption is found to be $90~\mu W$ excluding the RF module, which suggests that the sensor can also be powered without a battery via energy harvesting techniques using solar cell, vibration (piezo electricity), etc. In contrast, the power consumptions of the microprocessor-based digital sensors are at least two orders of magnitude higher than our ADB-based AJSCC circuit design.

\textbf{Our Approach:}
To overcome these limitations, we propose new baseband signal compression and processing in all-analog components to replace the ADC, digital chipset, and DAC, therefore achieving power-consumption reduction at the baseband components. 
This approach does all the processing in the analog domain (without the need for ADC, DAC and digital chipsets), and thus greatly reduces power consumption as well as circuit complexity compared with a digital-based sensor design. In addition, our structure retains some of the key functionalities of the digital system such as signal compression and multiplexing, thus allowing power-efficient transmission of the sensed signals from a large number of all-analog sensors. In particular, the signal-compression functionality is realized with innovative circuit designs that realize rectangular-type Analog Joint Source Channel Coding~(AJSCC), which is the source compression method proposed by Shannon in his 1949 seminal paper~\cite{Shannon49}. To the authors' best knowledge, ours is the first all-analog circuit realization of AJSCC, which replaces the digital processing components with a complete analog structure, resulting in low circuit complexity and low power consumption. In this article, we present two analog designs of the rectangular-type AJSCC circuit: i)~Design~1---circuit realized by parallel Voltage Controlled Voltage Source~(VCVS), which is our previous work~\cite{Zhao16}; ii)~Design~2---circuit realized by Analog Divider Blocks~(ADB), an improvement over Design~1 in terms of power and cost, which are critical requirements in high-density persistent wireless monitoring solutions. In this work, we are focusing on designing low-power and low-cost encoding circuits (based on rectangular Shannon Mapping) for transmission only. We assume the receiver/cluster head is a resource-rich mote with digital processing units and, as such, the decoding does not need to employ all-analog low-power circuits. Finally, even though we illustrate our design with temperature and humidity signals, it is generalizable to any two analog signals as our design does not pose any restriction on the type of signals as long as they are analog in nature.

\textbf{Choice of Rectangular-type AJSCC}:
This choice is motivated by the circuit realization advantage of the rectangular-type AJSCC over the alternate spiral type. Due to the spiral-type's unique structure, which is very difficult to fit into any of the existing analog circuits, its realizations are based on digital circuits. Specifically, the challenge in designing an all-analog spiral-type AJSCC circuit is to find an analog circuit that can realize the space-filling property of a spiral curve. The inputs of the circuit are two analog signals, and the output should be only one analog signal representing the accumulated length of the curve from the origin to the mapped point in such a way that, at the receiver, the two analog signals can be mapped back by the curve. As this is still an open-research analog design challenge, in this article we focus on the analog circuit realization of rectangular-type mapping.

\textbf{Sensor Multiplexing:}
The sensor nodes in the cluster are multiplexed by frequency division multiplexing, i.e., each node occupies a specific frequency band. The nodes are identified by the frequency band assigned at the initial design of the sensing system. Note that there are no digital packets and no digital synchronization designed in the system. The signals from the sensor nodes can be sent periodically with a null period to save transmission power. The receiver/cluster head samples the analog signal and recovers the compressed sensor signal at the frequency band assigned. The reason that the system does not need synchronization is that any sampled signal at a particular time instance contains the two compressed signals at that time instance, and the decoding process is the reverse mapping on the AJSCC curve. sensor multiplexing schemes are also being proposed by us based on frequency position modulation ~\cite{mass17}.

\textbf{Our Contributions:}
They can be summarized as follows.
\begin{itemize}

\item We propose an improved AJSCC circuit design by Analog Divider Blocks with tunable AJSCC-level spacing;

\item We verify the functions of the proposal with LTSpice simulations, and estimate its power consumption;

  \item We compare the proposal with previous design to show the lower power and circuit complexity advantages of the improved circuit.

\end{itemize}

\textbf{Outline:}
In Sect.~\ref{sec:ajscc_overview}, we present the principle of rectangular-type AJSCC and its existing circuit realizations.
In Sect.~\ref{sec:design1}, we review the parallel-VCVS-based Design~1.
In Sect.~\ref{sec:design2}, we introduce Design~2, which is based on Analog Divider Blocks, as an improvement over Design~1. 
Finally, in Sect.~\ref{sec:conc}, we draw the conclusions and discuss future directions.

\begin{figure}
\begin{center}
\includegraphics[width=3.2in]{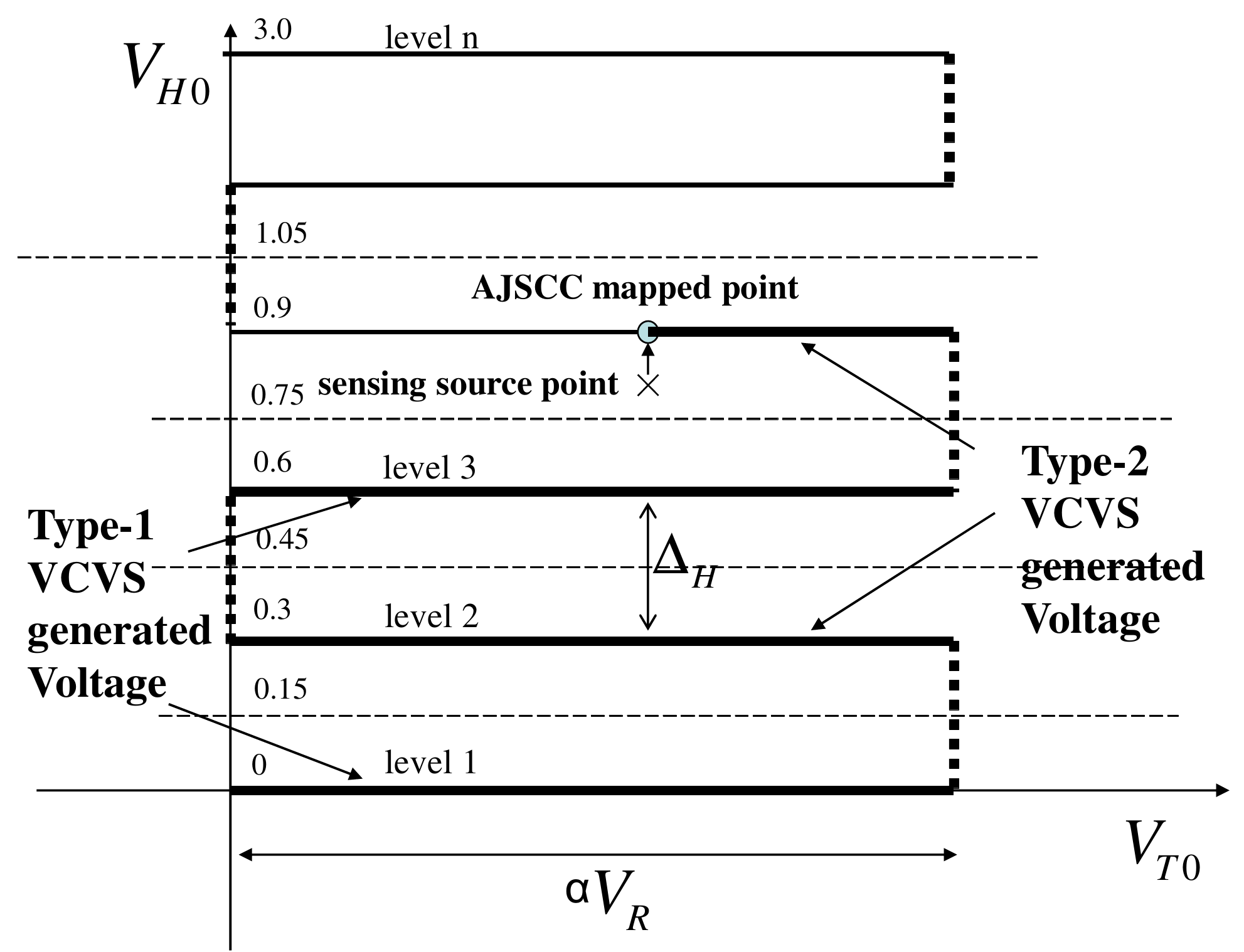}
\end{center}
\caption{\small{Shannon's Rectangular Mapping~\cite{Shannon49}. The sensed point is mapped to the point closest on the rectangular curve, and the curve accumulated length from the origin to the mapped point (in bold) is transmitted instead of the two values identifying the 2D sensed point. Here the numbers are the voltages used in our Spice simulations.}}\label{fig:rec_mapping}
\vspace{-0.15in}
\end{figure}

\section{AJSCC and Existing Circuit Realizations}\label{sec:ajscc_overview}
Analog Joint Source Channel Coding~(AJSCC)~\cite{Hekland05} can compress two or more sensor signals into one with controlled distortion while also being robust against wireless channel impairments. AJSCC adopts Shannon mapping as its encoding method~\cite{Fresnedo15}. Such mapping, in which the design of \emph{rectangular (parallel) lines} can be used for 2:1 compression (Fig.~\ref{fig:rec_mapping}), was first introduced by Shannon in his seminal 1949 paper~\cite{Shannon49}. Later work has extended this mapping to a \emph{spiral type} as well as to N:1 mapping~\cite{Brante13}. AJSCC achieves optimal performance in rate-distortion ratio, whereas to achieve such optimality using \emph{separate} source and channel coding complex encoding/decoding and long block-length codes would be required, causing delays and energy inefficiencies. Shannon mapping has the two-fold property of (1)~compressing the sources (by means of N:1 mapping) and (2)~being robust to wireless channel distortions as the noise only introduces errors along the parallel lines (or the spiral curve). 

AJSCC requires simple compression and coding at the transmitter, and low-complexity decoding at the receiver. In rectangular-type mapping, to compress the source signals (``sensing source point"), say Humidity and Temperature voltages ($V_H$, $V_T$), the point on the space-filling curve with minimum Euclidean distance from the source point is chosen (``AJSCC mapped point"), as illustrated in Fig.~\ref{fig:rec_mapping}, via a simple projection on the curve. The transmitted signal is then the ``accumulated length'' of the lines from the origin to the mapped point, where the error introduced by the mapping is controlled by the spacing $\Delta_H$ between lines. At the receiver, the received signal of the accumulated length with noise is mapped to one point on this rectangular curve, and the two original points can be recovered. This is called the rectangular-type AJSCC. In this article, we discuss power-efficient and low-complexity analog circuit realizations of this type of AJSCC, for two-to-one (2:1) analog signal compression in wireless sensors.

Existing AJSCC solutions use \emph{digital hardware}, and are power demanding and of high circuit complexity. A Software-Defined Radio~(SDR) system to realize AJSCC mapping has been reported in~\cite{Garcia11}. The mapping was also recently implemented in an optical digital communication system in~\cite{Romero14}. Shannon-mapping encoding was adopted in~\cite{stopler14} for a digital video transmission. All existing circuits adopt the structure of ADC, followed by digital processing chipset, then DAC to RF. These designs have the disadvantages of high power consumption and circuit complexity. No existing works have implemented AJSCC using an all-analog, low-complexity circuit design. In the following two sections, we describe two AJSCC circuits we have designed and realized using \emph{all-analog} components.

\section{Review of Design with Parallel VCVS (Design~1)}\label{sec:design1}
\begin{figure}
\begin{center}
\includegraphics[width=3.6in]{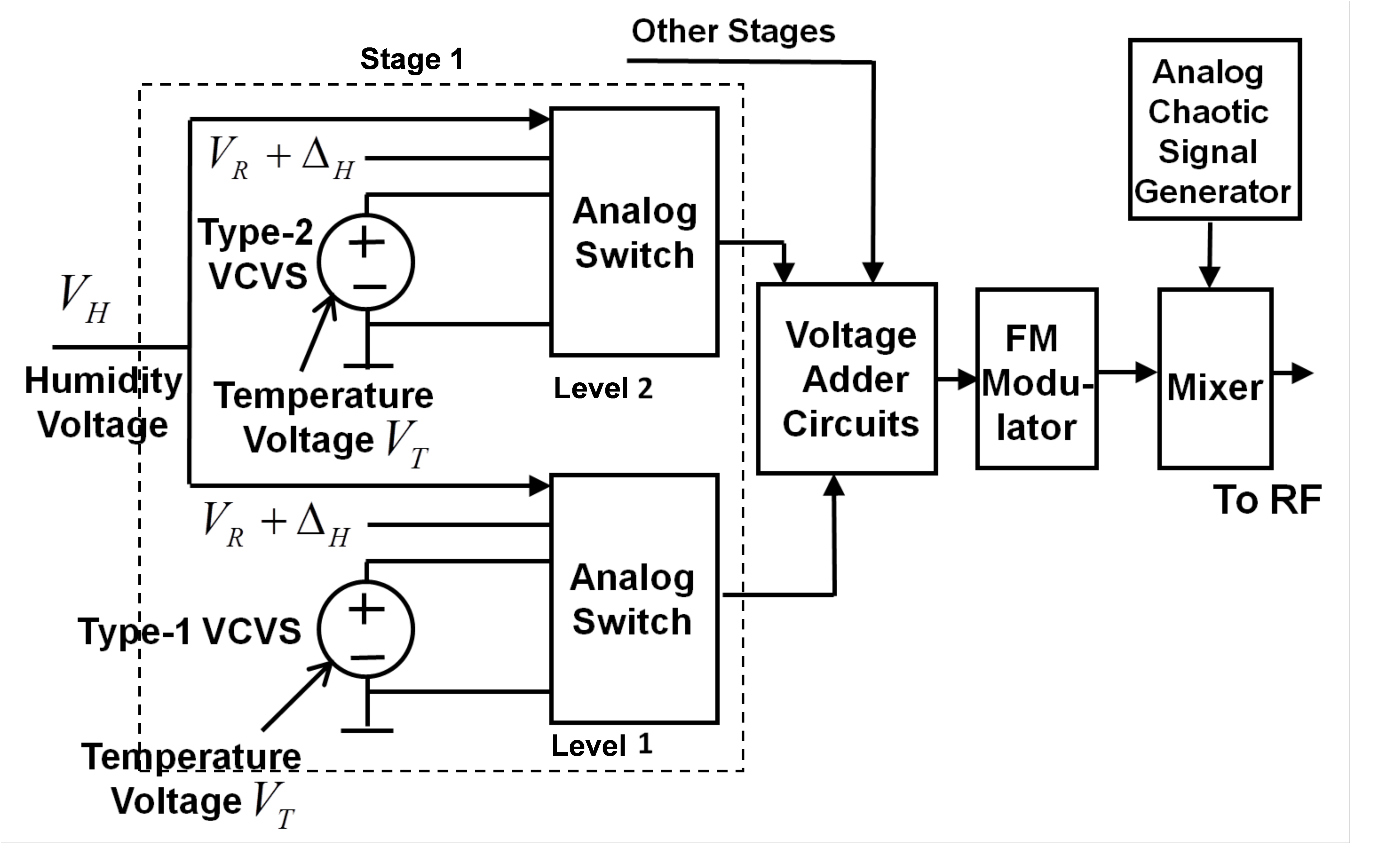}
\end{center}
\caption{Original analog circuit of parallel-VCVS design for Shannon's rectangular mapping (Design~1); for simplicity only the first stage is depicted.}\label{fig:prop_ckt}
\vspace{-0.15in}
\end{figure}
We had proposed a circuit realization with parallel VCVS in~\cite{Zhao16}. Here, we review the key design concepts of this parallel-VCVS design so as to prepare the reader for the new and improved AJSCC circuit design, Design~2, in the next section.

\textbf{Design Overview:}
Our original analog circuit is depicted in Fig.~\ref{fig:prop_ckt}. In our all-analog implementation by parallel VCVS, odd-level voltages are generated using Type-1 Voltage Controlled Voltage Sources~(VCVS), whereas even-level voltages are generated using Type-2 VCVS. $V_H$ and $V_T$ are the humidity and temperature variable voltages, and $V_{H0}$ and $V_{T0}$ are the humidity and temperature variable voltages after range adjustment (i.e., offset removal). $\Delta_H$ is the spacing between the levels and $V_R$ is the voltage of one parallel line in the mapping curve, which is a constant voltage value. As such, it is chosen to be proportional to the maximum value of $V_{T0}$ and vice-versa, as shown in Fig.~\ref{fig:rec_mapping}.

The VCVS accepts voltages $V_{T0}$ and $V_{H0}$ as inputs, and outputs a voltage that is a function of the input voltage. By observing the rectangular-mapping curve (see Fig.~\ref{fig:rec_mapping}), there are two types of output increments. In the first (which we call Type~1), the output $V_O$ increases linearly with increasing $V_{T0}$, i.e., $V_{O,Type1} \propto V_T$. This happens when we traverse the curve from left to right on odd-numbered lines to reach the mapped point. In the second (Type~2), the output decreases linearly with increasing $V_T$, i.e., $V_{O,Type2}=V_R-V_{O,Type1}$, which corresponds to traversing the curve from right to left on even-numbered lines to reach the mapped point. We realize these two types of outputs using two types of VCVS, where the overall mapping output is the voltage summation of the activated VCVS blocks, which represents the accumulated length of the curve from the origin to the mapped point. Note that $V_{H0}$ controls which VCVS levels/lines are activated, while $V_{T0}$ controls the VCVS output. A stack of analog switches is used to implement the former and also to control the output of all levels. The number of switches is determined by the mapping resolution sought ($\Delta_H$).

Each parallel line is defined as a \textit{level} and is numbered from the bottom upward. The first and second levels are combined to be defined as the first \textit{stage}. The second stage will include the third and fourth levels and so on. The first stage is shown in Fig.~\ref{fig:prop_ckt}, where the output of each level is controlled by the corresponding analog switch. Each switch has three inputs: $V_R+\Delta_H$, $GND$, and the output of either a Type-1 or Type-2 VCVS. Odd- or even-numbered switches are connected to Type-1 or Type-2 outputs, respectively. Note that the number of activated switches and the activation voltage of each switch are determined by considering both $V_{H0}$ and the chosen resolution $\Delta_H$. The switching logic is explained as follows: each switch is activated only if $V_H$ is greater than a certain value based on the switch's level in the stack. If not activated, the output of the switch is connected to $GND$; if activated, the output is equal to the VCVS output within a certain range ($\Delta_H$) of $V_{H0}$ above the activation voltage. Once $V_{H0}$ exceeds this range, the switch outputs its saturation voltage of $V_R+\Delta_H$. Even though $\Delta_H$ is used to calculate the length of the curve from the origin to the mapped point, it does not provide any additional information. Hence the saturation voltage can just be $V_R$, instead of $V_R+\Delta_H$.
Since neglecting this part of the length of the curve (i.e., $\Delta_H$) will not affect the accuracy of the decoded signal and can simplify the mapping, in both simulation and hardware implementation of Design~1 this $\Delta_H$ gap has been neglected.

\begin{figure*}[ht]
        \centering
           \begin{subfigure}[b]{0.31\textwidth}
        		\centering
        		 \includegraphics[width=1\textwidth,height=1.8in]{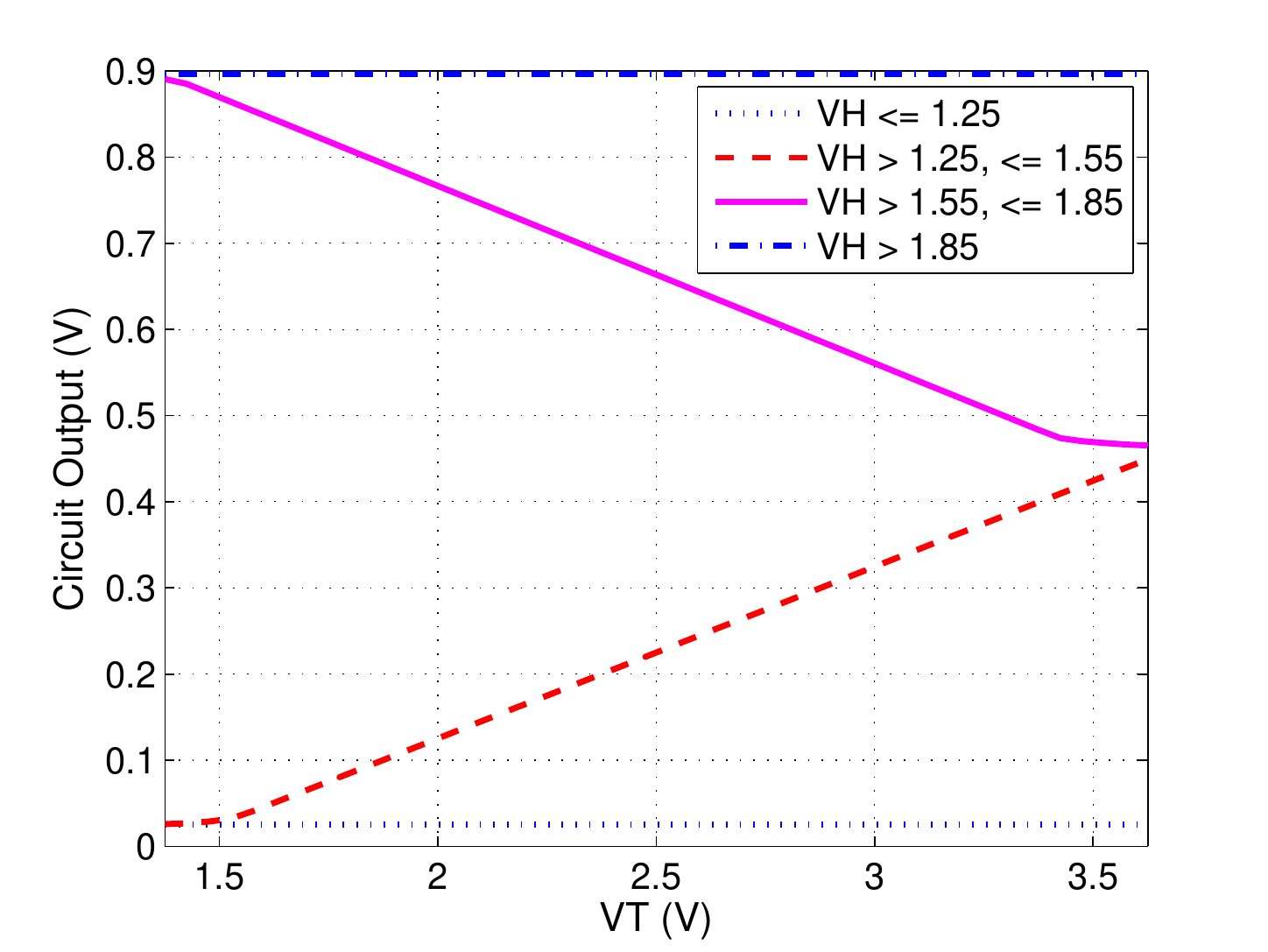}
        		\caption{}
        		\label{fig:spicebb_results}
        	\end{subfigure}
~
        \begin{subfigure}[b]{0.31\textwidth}
            \centering
            \includegraphics[width=1\textwidth,height=1.8in]{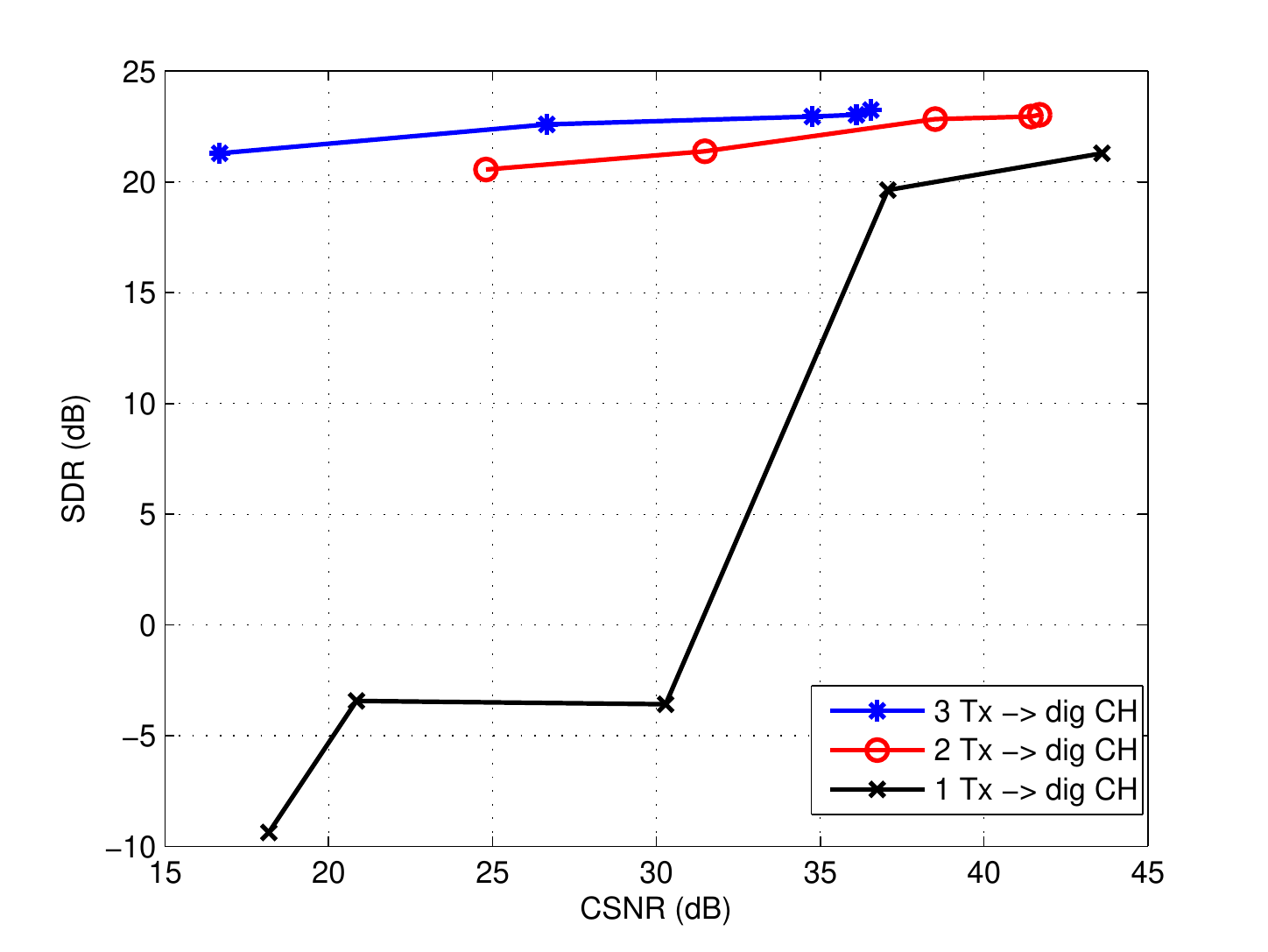}
            \caption{}
            \label{fig:csnr_sdr_combined}
        \end{subfigure}
~
        \begin{subfigure}[b]{0.31\textwidth}
            \centering
            \includegraphics[width=1\textwidth,height=1.8in]{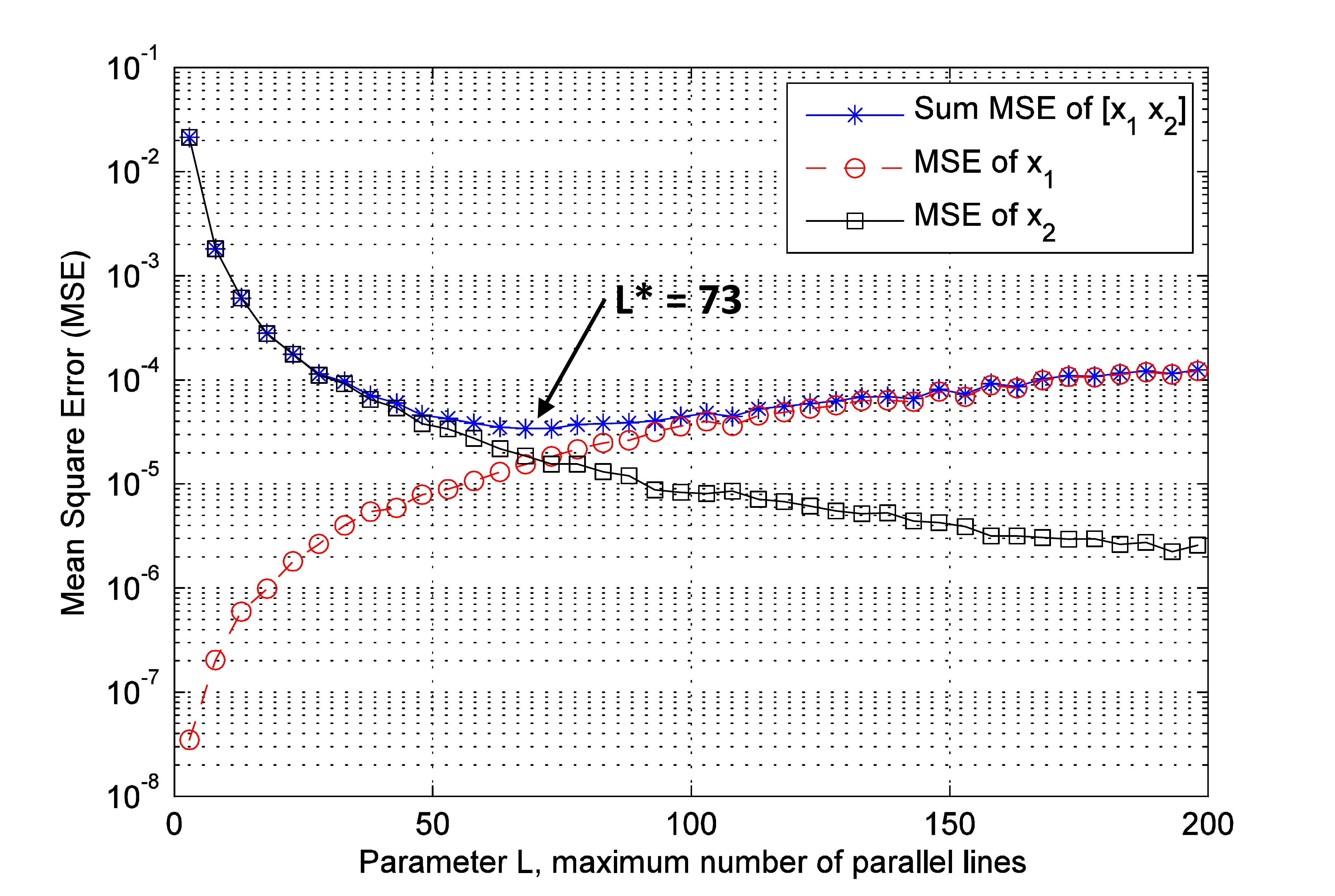}
            \caption{}
            \label{fig:mse-vs-levels}
        \end{subfigure}
        \caption{(a)~AJSCC output of the \emph{second stage}---notice the four different regions: off, linear (Type-1 VCVS), linear (Type-2 VCVS), and saturation; (b)~Measured SDR-vs-CSNR performance for one, two, and three Tier-1 all-analog sensors communicating with a Tier-2 digital Cluster Head~(CH).;
(c)~Mean Square Error~(MSE) vs. number of levels ($n$=$L$) for SNR=-$20~\rm{dB}$, with optimum $n$=$73$; similar trends are observed for SNR=-$10$, $0~\rm{dB}$.}
        \vspace{-0.15in}
\end{figure*}

The AJSCC mapping circuit has been simulated in LTSpice using manufacturer-provided Spice models to capture real performance; Figure~\ref{fig:spicebb_results} shows the output of the second level over its entire mappable range of $V_T$ and $V_H$. It can be seen that the output is almost zero when $V_H$ is below $1.25~\mathrm{V}$ and that, when $1.25 < V_H \le 1.55$, Type~1 mainly drives the output (with zero contribution from Type~2); conversely, when $1.55 < V_H \le 1.85$, Type~2 mainly drives the output (with $V_R$ contribution from Type~1); finally, when $V_H > 1.85$, the output is $2V_R$ (i.e., $V_R$ from each VCVS). The detailed information on the LTSpice simulation results can be found in~\cite{Zhao16}.

\begin{figure}
\begin{center}
\includegraphics[width=2.5in,height=1.8in]{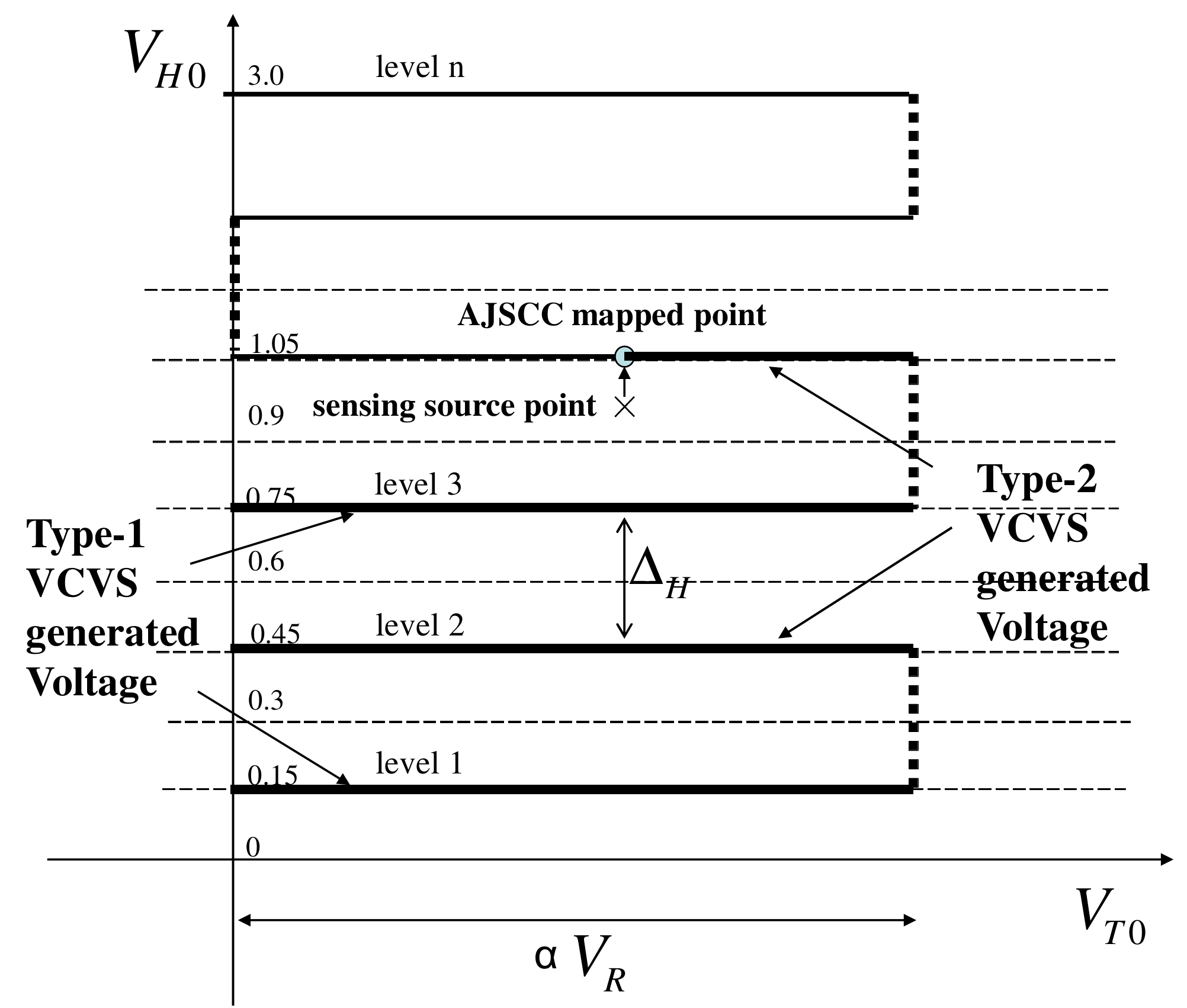}
\end{center}
\caption{Modified Shannon mapping with half $\Delta_H$ offset to maximize the efficiency of the new circuit.
}\label{fig:new_rec_map}
\end{figure}

\textbf{SDR vs. CSNR Results:} 
The proposed system has been implemented in a prototype composed of multiple sensor nodes and one cluster head node. The sensor nodes are multiplexed by frequency multiplexing on adjacent and separated channels. The two sensing signals are first compressed by the AJSCC circuit, then modulated using frequency modulation by a Radio Frequency Integrated Circuit (RFIC), then sent to the antenna, all realized on a PCB. At the receiver, the signal is digitally sampled, and then frequency-demodulated and AJSCC-decoded to recover the original signals. Figure~\ref{fig:csnr_sdr_combined} compares the Signal-to-Distortion Ratio versus Channel Signal-to-Noise Ratio (SDR-vs-CSNR) performance of one, two, and three analog sensors communicating to a digital cluster head using different channels showing the effect of receiver diversity. We clearly observe that the three-sensor case has better performance than the two-sensor case, which in turn is far better than the single-sensor case thanks to channel diversity at the receiver. For further details, interested readers can refer to~\cite{wons3tier2017}.

\begin{figure}
\begin{center}
\includegraphics[width=2.5in,height=1.8in]{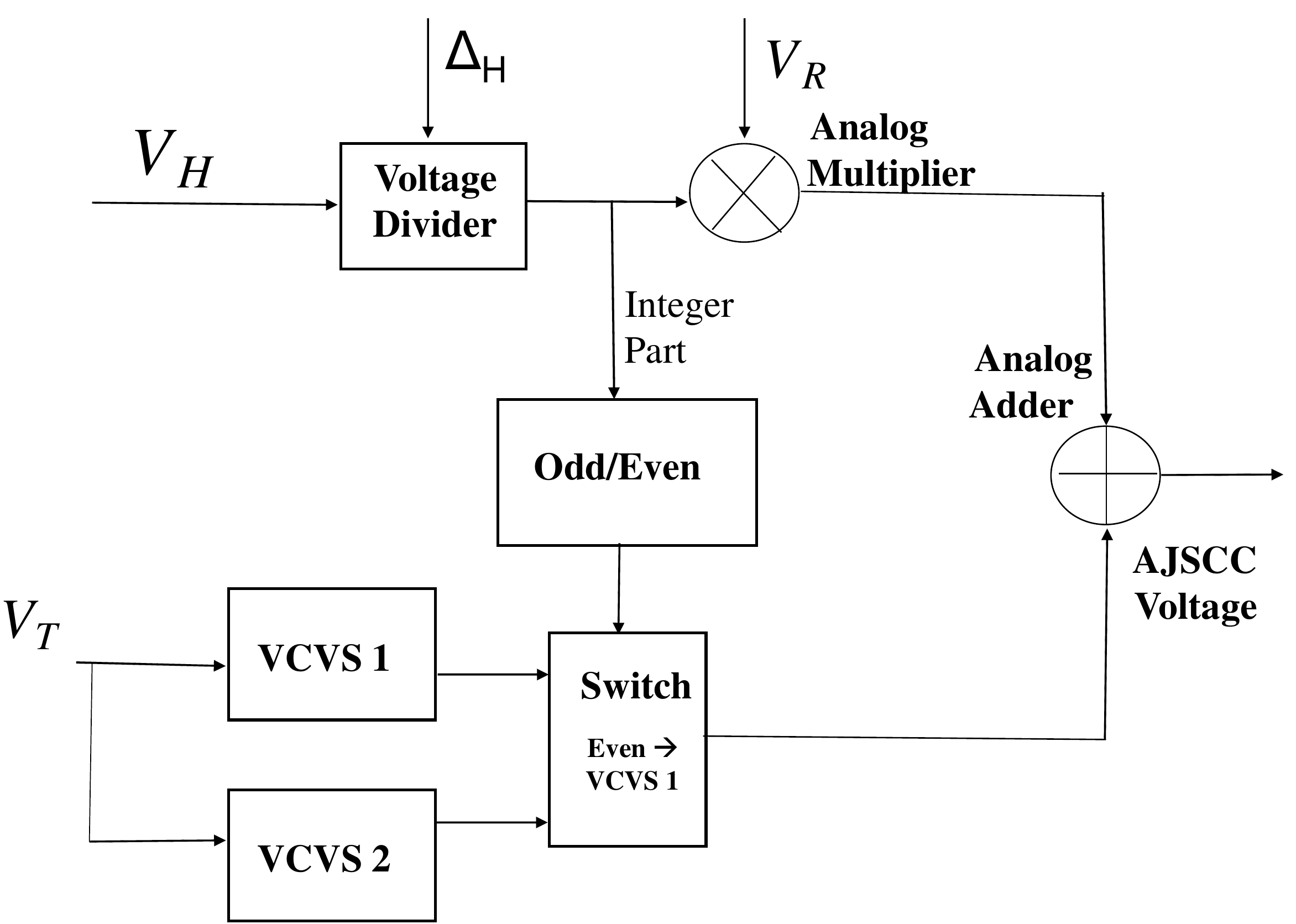}
\end{center}
\caption{High-level overview of Design~2, where Voltage Divider and Odd/Even blocks are realized using the proposed Multi-stage Analog Divider circuit.}\label{fig:design_2}
\end{figure}

\textbf{Optimum Number of AJSCC Levels:}
We note an interesting tradeoff on the number of AJSCC levels used. With increasing number of levels, $n$, the Mean Square Error~(MSE) of $V_H$ drops as the spacing between the lines ($\Delta_H$) reduces. However, since the total encoded voltage cannot be arbitrarily high (as it is limited by the supply voltage), the voltage representing $V_T$ will be smaller, which will increase the MSE for $V_T$. On the other hand, with decreasing $n$, the quantization error in $V_H$ increases, leading to high MSE for $V_H$. We have thoroughly studied this tradeoff via MATLAB simulations and found an optimal $n$ (=$L$), as in Fig.~\ref{fig:mse-vs-levels} (for further details refer to Sect.~III-C of~\cite{wons3tier2017}).
We note that the optimum number of levels (for minimum MSE) is a large number (around 73), which is difficult to implement using Design~1 (in terms of both power and cost). We observe from Fig.~\ref{fig:mse-vs-levels} that in increasing from 11 levels (as in Design~1) to 64 levels (feasible with Design~2), the signal recovery MSE is reduced dramatically by an order of magnitude for the three SNR values evaluated. This presents compelling motivation to develop Design~2, which can support a large number of levels and achieve optimality.

\begin{figure*}
\begin{center}
\includegraphics[width=0.81\textwidth]{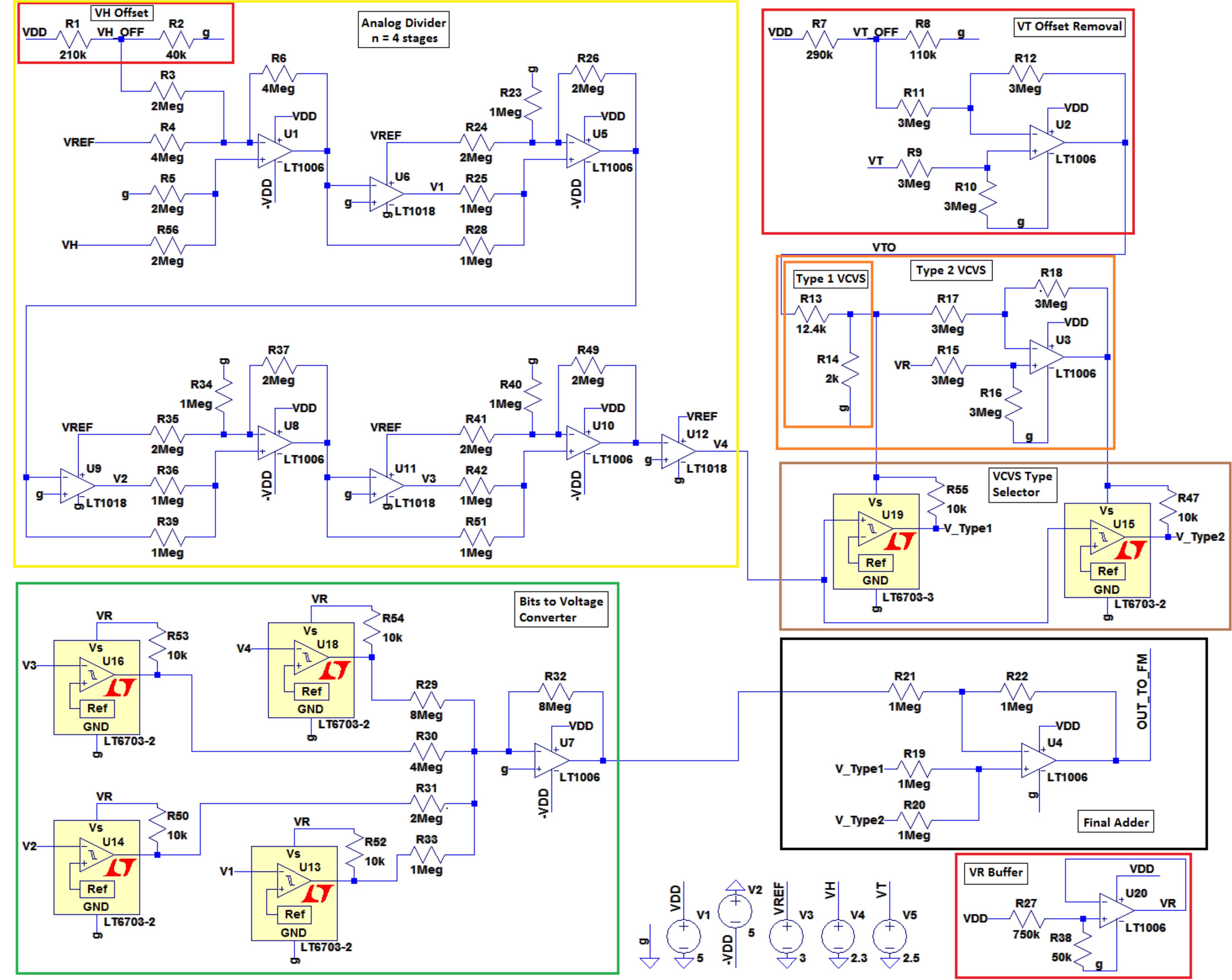}
\end{center}
\caption{LTSpice schematic of the proposed circuit with parameter $k=4$.}\label{fig:prop_ckt_2}
\vspace{-0.15in}
\end{figure*}

\section{Improved Design with Multistage Analog Divider Block (Design~2)}\label{sec:design2}
The circuit in our previous work grows linearly in complexity with the number of levels, $n$ (but exponentially with the number of components used), which is not desirable for low-power sensing at a large number of levels. Thus, developing a method with better than $O(n)$ linear complexity would lower power consumption for large n. Hence, in Design~2, we aim at achieving $O(ln(n))$ complexity (or linear in the number of components used). In the following content, we firstly describe the design methodology of Design~2; then present the circuit design and LTSpice simulation results. The power consumption of Design~2 is elaborated and, finally, power and cost comparisons with Design~1 are presented.

\textbf{Design Methodology:}
In this improved second design, we propose to shift the parallel lines in Fig.~\ref{fig:rec_mapping} upward by half $\Delta_H$, as shown in Fig.~\ref{fig:new_rec_map} for the implementation of the proposed design. The new mapping offsets all of the levels by $V_{H0}=\frac{\Delta_{H}}{2}$ and removes the highest level. This mapping makes the circuit more efficient as it removes the asymmetry in the first level (i.e., the mapping range from $V_H = -0.15~\text{to}~0~\rm{V}$ is not available in the example shown in Fig.~\ref{fig:rec_mapping} unlike other levels, as we consider only positive sensor values). With this background, the improved circuit to realize the AJSCC mapping with arbitrary $\Delta_H$ is shown in Fig.~\ref{fig:design_2}. The input voltage $V_H$ is firstly divided by a tunable voltage of $\Delta_H$---the desired spacing between the ASJCC levels. The voltage divider outputs the integer part of the division, and this integer is input to two blocks, the first is the Odd/Even block and the second is the Analog Multiplier block, where the former will determine if the input integer is odd or even, whereas the latter will multiply the integer with voltage $V_R$. In the Odd/Even block, an even-integer input controls the switch to accept the output of VCVS~1, the type-1 VCVS voltage (as indicated in Fig.~\ref{fig:design_2}); conversely, for an odd-integer input, the switch will accept the output of VCVS~2, the type-2 VCVS voltage. The voltage $V_T$ is input to the two types of VCVS to generate their respective outputs (proportional and inversely-proportional, respectively), which are fed to the switch. The analog multiplier output and the switch output are added by an analog adder to produce the final AJSCC signal. The Analog division and Odd/Even block are combinedly realized using multistage Analog Divider Block~(ADB), as explained below.

\begin{figure*}[ht]
        \centering
            \begin{subfigure}[b]{0.31\textwidth}
         		\centering
        		\includegraphics[width=1\textwidth]{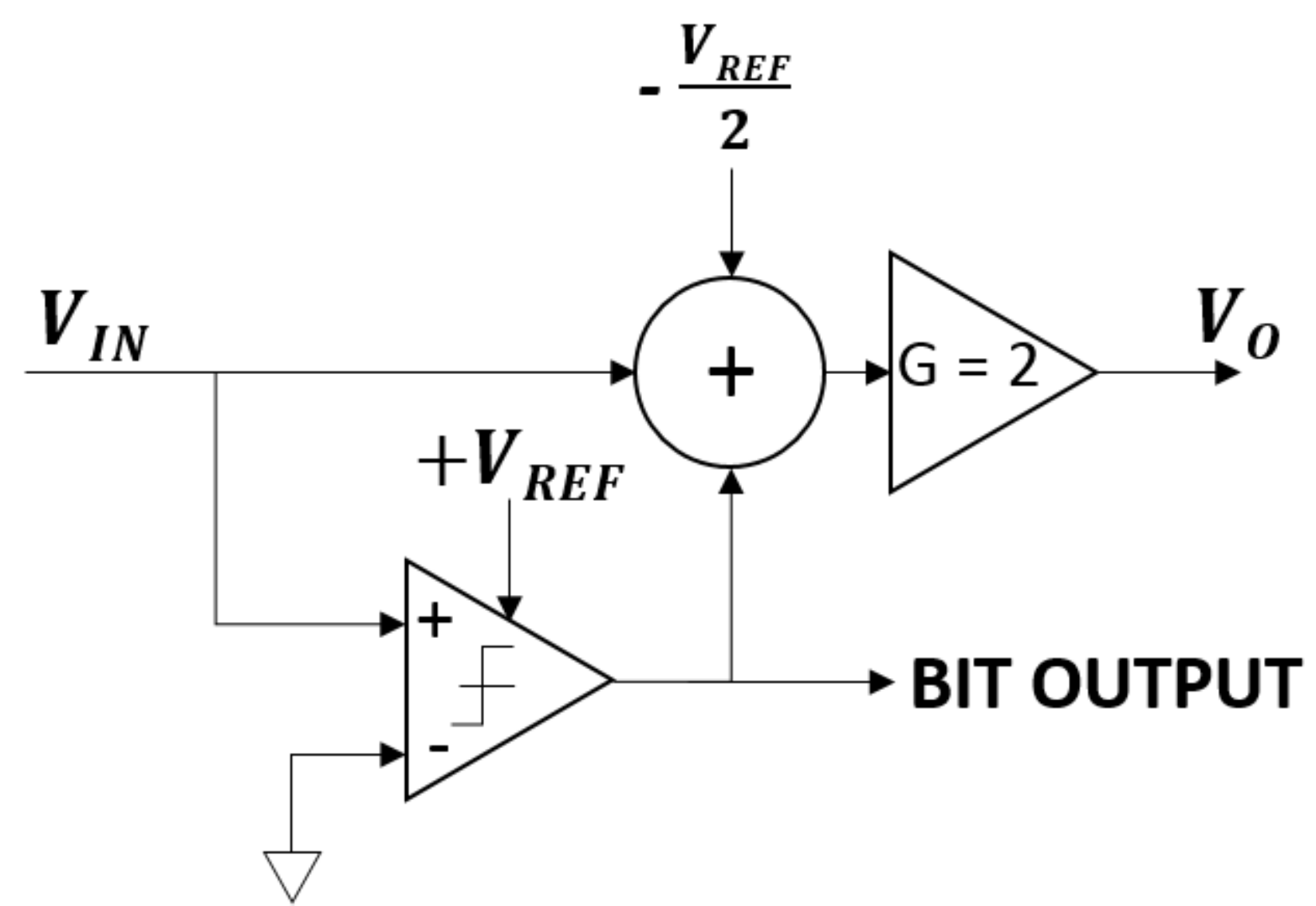}
         		\caption{}
         		\label{fig:ADB1}
         	\end{subfigure}
~
        \begin{subfigure}[b]{0.31\textwidth}
            \centering
            \includegraphics[width=1\textwidth]{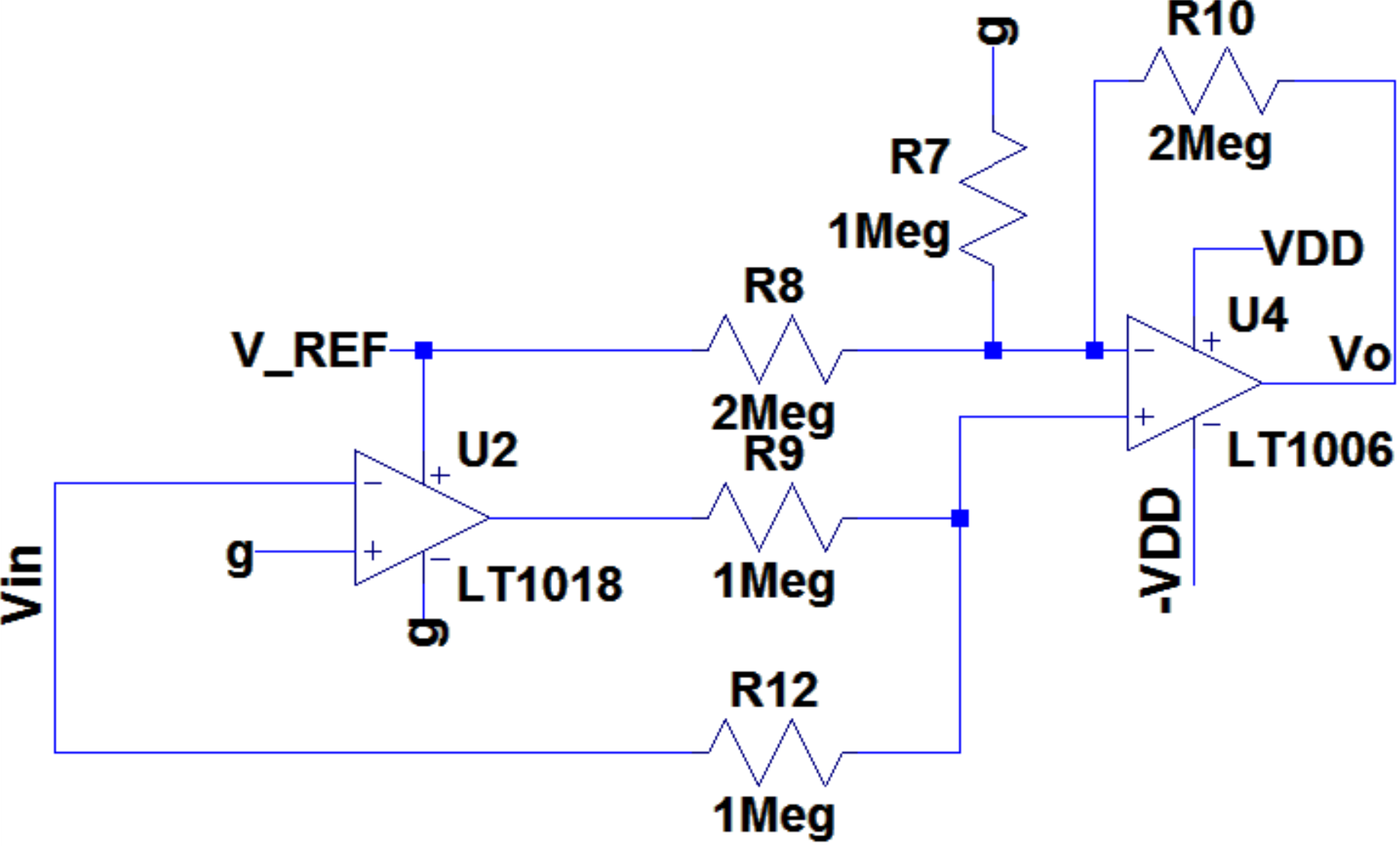}
            \caption{}
            \label{fig:Modulus}
        \end{subfigure}
~
         \begin{subfigure}[b]{0.31\textwidth}
             \centering
             \includegraphics[width=1\textwidth]{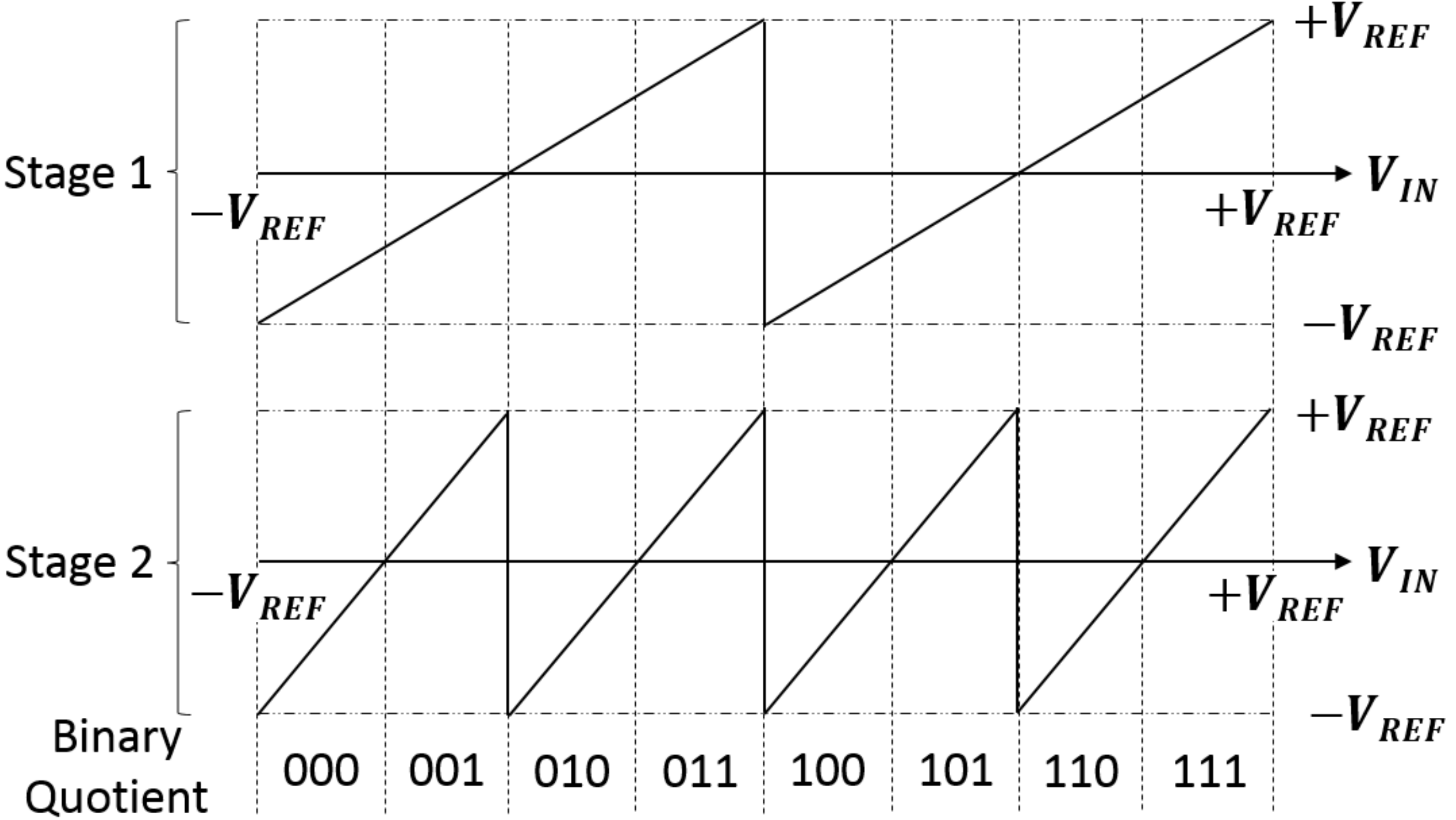}
             \caption{}
             \label{fig:ADB2}
         \end{subfigure}
         \caption{(a)~High level view of a single stage analog divider with one analog divider block (ADB), which performs modulo by $V_{REF}$ within range [${-V}_{REF}$,${V}_{REF}$]; (b)~LTSpice schematic of (a); (c)~Plot showing how a two-stage analog divider can create a binary representation of the quotient---the output of stage~1 determines the first bit from the input voltage, $V_{in}$; then, the stage~1 output is fed to the stage~2 to determine the second bit~\cite{foldingadc}.}
  \vspace{-0.15in}
\end{figure*}

\textbf{Circuit Implementation:} In the following multi-stage ADB circuit, the term ``stage'' refers to an analog divider block, and parameter $k$ denotes the number of such stages. (Note that usage of ``stage'' here is different than in Design 1.) A circuit with $k$ stages results in $n=2^{k}$ levels, and Fig.~\ref{fig:prop_ckt_2} shows the proposed circuit for $k=4$, corresponding to 16 levels. This design uses VCVS output to generate each level as in the previous circuit, but differs in how a sensed point is mapped to a particular level and as well as on what VCVS type is chosen for that level. Instead of using comparators between every individual level, the proposed circuit identifies the appropriate level number by dividing $V_H$ by $\Delta_H$ using the multi-stage ADB. The resulting quotient is output in bits and the smallest bit is used to determine whether to use Type-1 or Type-2 VCVS. The quotient bits are converted back to voltage using comparators and a summing amplifier. Note that this is done solely in the analog domain, without the use of any power-hungry ADCs. The circuit can be configured to have any number of stages, $k$, depending on the resolution requirements of the application.
In the rectangular mapping used for this circuit, each subsequent level uses the opposite VCVS from the previous level. The result is that all odd-numbered levels use the Type-1 VCVS while even-numbered levels use the Type-2 VCVS. Thus, the problem of selecting which VCVS to use reduces to determining if the level is odd or even, removing the need to have the linear cascade design as in Design~1.

\begin{figure*}[ht]
        \centering
           \begin{subfigure}[b]{0.32\textwidth}
        		\centering
        		 \includegraphics[width=1\textwidth]{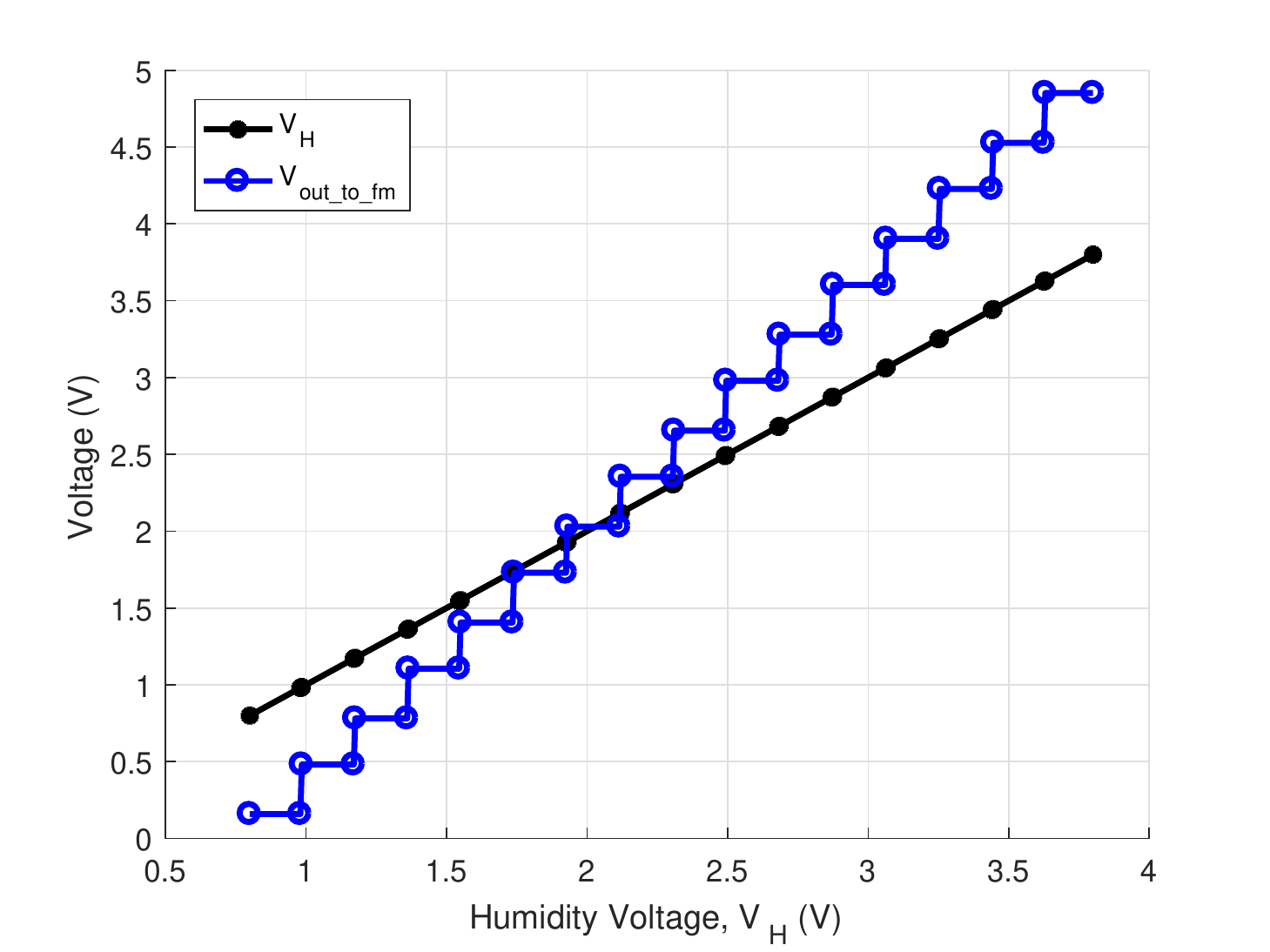}
        		\caption{}
        		\label{fig:LevelsPlot}
        	\end{subfigure}
~
        \begin{subfigure}[b]{0.32\textwidth}
            \centering
            \includegraphics[width=1\textwidth]{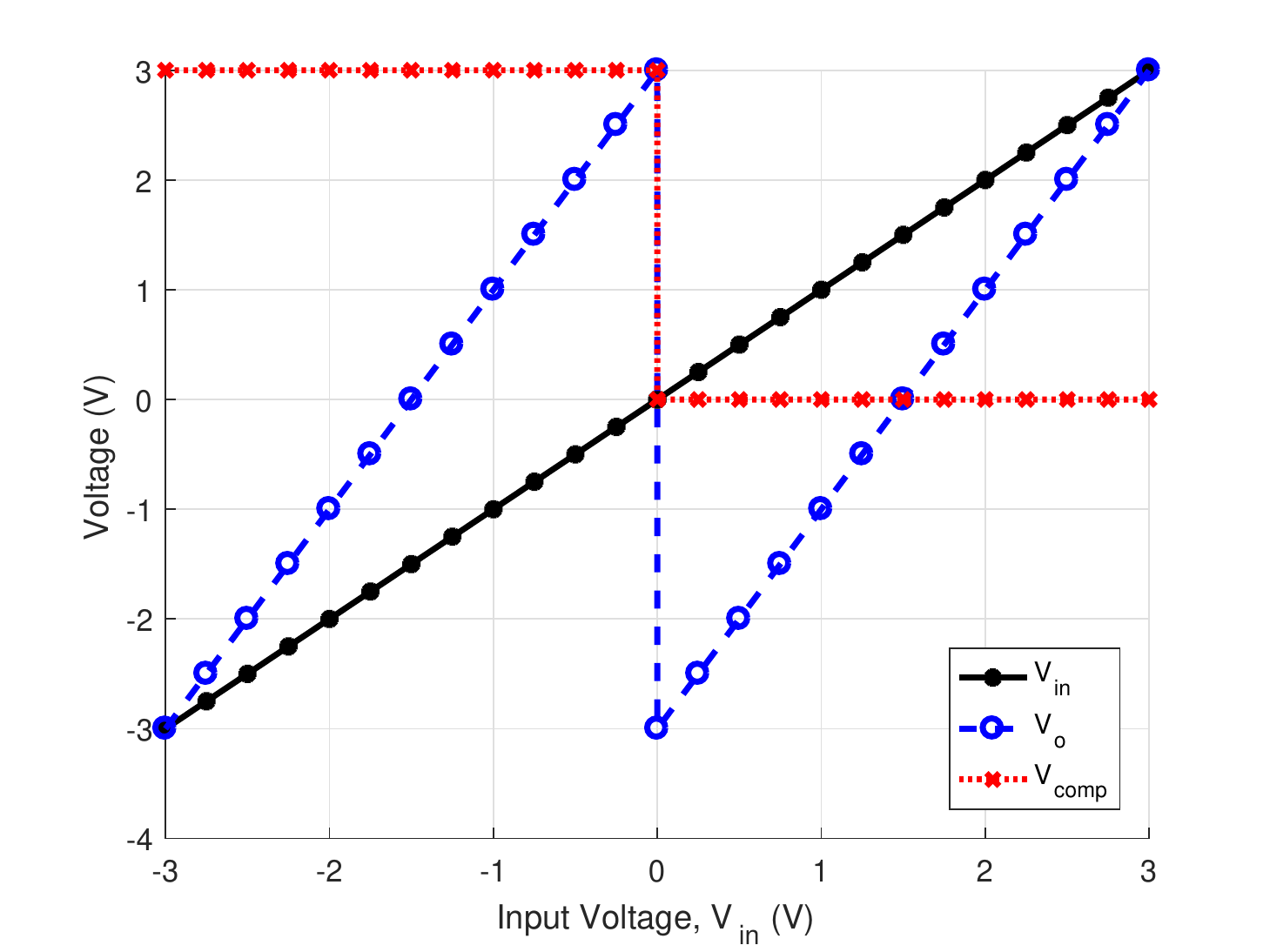}
            \caption{}
            \label{fig:ModulusPlot}
        \end{subfigure}
~
        \begin{subfigure}[b]{0.32\textwidth}
            \centering
            \includegraphics[width=1\textwidth]{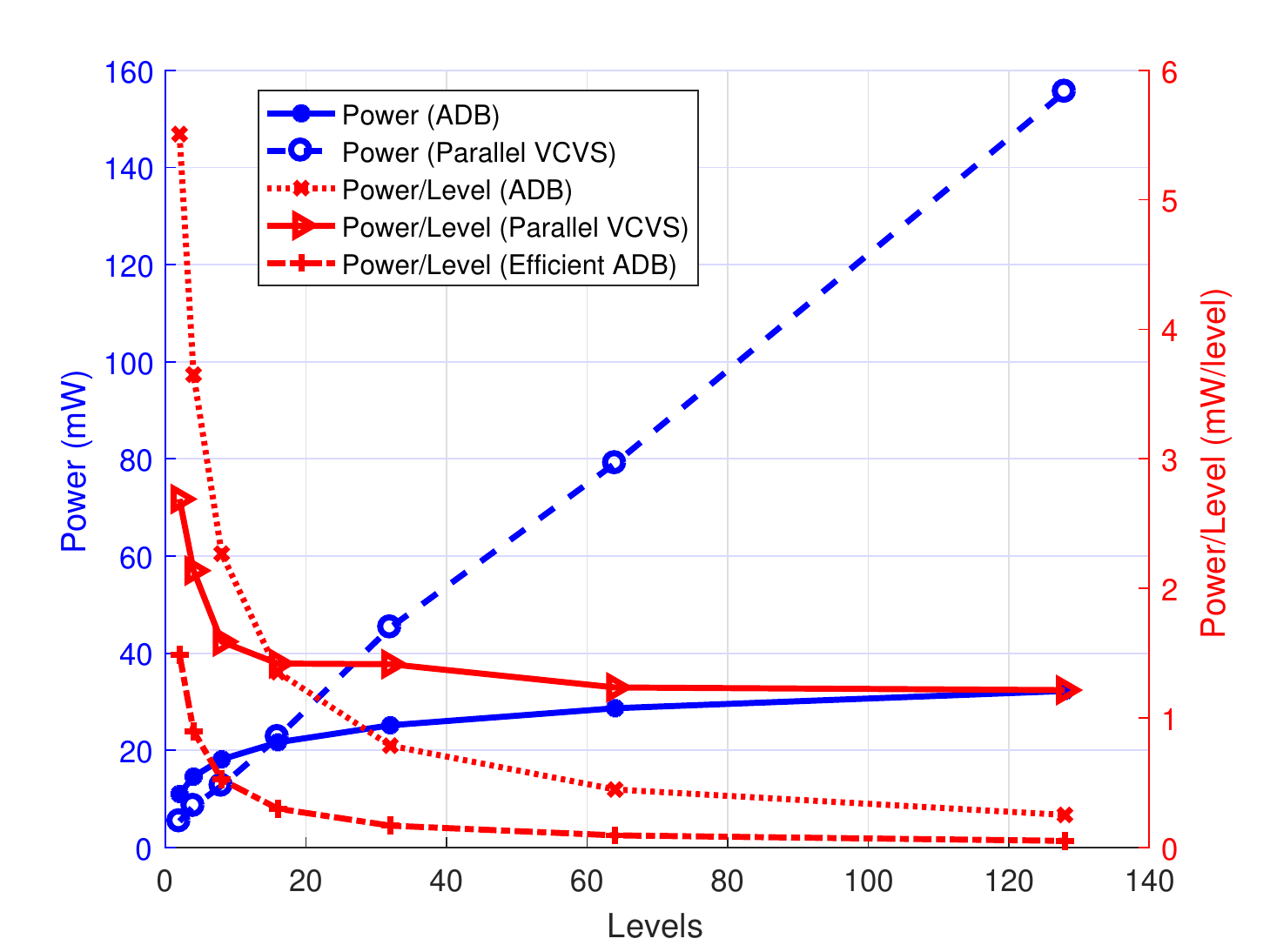}
            \caption{}
            \label{fig:PowerPlot}
        \end{subfigure}
        \caption{\label{fig:3dmap2} (a)~AJSCC encoded output of 16-level analog divider circuit with varying $V_{H}$ and $V_{T} = 2.5~\mathrm{V}$ and $V_{REF}=3~\rm{V}$; (b)~Output of one-stage analog divider circuit with $V_{in}$ ranging from $-3$ to $3~\mathrm{V}$, with $V_{REF}=3~\mathrm{V}$; (c)~Power (blue, left y-axis) and Power-per-level (red, right y-axis) consumed by the ADB design, parallel VCVS design, and ADB design with power efficient COTS components.}
        \vspace{-0.15in}
\end{figure*}

A simple way to determine parity is to use the modulo operation.
For example, to determine if an integer $M$ is even or odd, $M$-mod-$2$ results in either 0 if \textit{M} is even, or 1 if \textit{M} is odd. If $M$ is a real number instead of an integer, it cannot directly be classified as even or odd. However, the parity of $floor(M)$ can still be found from $M$-mod-$2$, whose result will be a real number ranging in $[0,2)$. If the result is within the range $[0,1)$, then $floor(M)$ is even; if it is within $[1,2)$, $floor(M)$ is odd. To apply this concept to the circuit, we calculate $V_{H0}$ mod by ${\Delta}_H$. If the result of this expression falls within the range $[0,\frac{{\Delta_H}}{2})$, then it belongs to an odd level, whereas if it is within $[\frac{{\Delta_H}}{2},{\Delta}_H)$ it belongs to an even level.
One way to perform the modulo operation in analog is to use a multi-stage analog divider~\cite{foldingadc}. The analog divider can find modulo of one voltage by a second voltage within a predefined range. Figure~\ref{fig:ADB1} shows one stage of the analog divider, which performs $V_{in}$ mod $V_{REF}$ where the input voltage, $V_{in}$ falls within $[-V_{REF},V_{REF}]$ and $V_{REF}$ is a reference voltage in the modulo circuit which is chosen to determine both the modulo divisor and the input voltage range. Figure~\ref{fig:Modulus} shows this stage implemented in LTSpice. Cascading $k$ such stages yields $V_{in}$ modulo by $\frac{V_{REF}}{2^{k}}$, with the same input range $[-V_{REF},V_{REF}]$, and $\Delta_H = \frac{V_{REF}}{2^{k}}$. In this design, to create the discontinuity around the y-axis, a comparator checks the input voltage and outputs an offset of $V_{REF}$ if the input is negative. This offset is added to the input voltage at a summing amplifier, which also adds another offset of $\frac{-V_{REF}}{2}$ to center the output about the x-axis. This is useful since the operating region of the Op Amp used in the summing amplifier is also centered around the x-axis. The sum is multiplied by $2$ to get the final output, called the residue of the stage. Eq.~\eqref{eq1} shows the output $V_o$ of the stage,
\begin{equation}
V_o =
\begin{cases}
2\times (V_{in}-\frac{V_{REF}}{2} + V_{REF}) & V_{in} <0, \cr 
2\times (V_{in}-\frac{V_{REF}}{2}) & V_{in}\geq0.
\end{cases}
\label{eq1}
\end{equation}

\begin{table}[t]\scriptsize
 \caption{$V_{REF}$, $\Delta_H$, and $V_R$ for up to $k=8$ analog divider stages; all voltages are in Volts.}\label{tab:levels}
\begin{center}
 \begin{tabular}{|p{0.2cm}|p{0.6cm}|p{0.8cm}|p{0.8cm}|p{0.6cm}|p{0.8cm}|p{0.8cm}|p{0.8cm}|}
 \hline
 \textbf{k} & \textbf{Max $n$} & \textbf{Min $V_{REF}$} & \textbf{Min $\Delta_H$} & \textbf{Min $n$} & \textbf{Max $V_{REF}$} & \textbf{Max $\Delta_H$} & \textbf{$V_{R}$}\\
 \hline\hline
 1 & 2 & 6 & 3 & - & - & - & 2.5\\
 \hline
 2 & 4 & 4 & 1 & 4 & 4 & 1 & 1.25\\
 \hline
 3 & 8 & 3.429 & 0.429 & 6 & 4.8 & 0.6 & 0.625\\
 \hline
 4 & 16 & 3.2 & 0.2 & 11 & 4.8 & 0.3 & 0.3125\\
 \hline
 5 & 32 & 3.097 & 0.097 & 21 & 4.8 & 0.15 & 0.1563\\
 \hline
 6 & 64 & 3.048 & 0.048 & 40 & 4.923 & 0.077 & 0.0781\\
 \hline
 7 & 128 & 3.024 & 0.024 & 78 & 4.987 & 0.039 & 0.0391\\
 \hline
 8 & 256 & 3.012 & 0.012 & 155 & 4.987 & 0.019 & 0.0195\\
 \hline
\end{tabular}
\vspace{-0.1in}
\end{center}
\end{table}

A cascade of analog divider stages can generate the binary equivalent of the quotient, as shown in Fig.~\ref{fig:ADB2}. If the output voltage for a stage is positive, this represents a 1 bit; whereas, if the output voltage is negative, it represents a 0 bit. The comparators are used to check whether the previous output is positive or negative by outputting low if it is a 1 bit and high if it is a 0 bit. Cascading $k$ stages gives a string of $k$ bits, with the most significant bit at the first stage in the cascade and the least significant bit at the final stage. The value of this binary bit code corresponds to the mapped level in the rectangular mapping. Cascading $k$ stages results in up to $2^k$ levels, with $\Delta_H=\frac{V_{REF}}{2^k}$. Table~\ref{tab:levels} shows how the number of levels and $\Delta_H$ scale with $k$. For each $k$ (i.e., corresponding hardware), it shows the minimum and maximum values of levels, $V_{REF}$, and $\Delta_H$ supported by the circuit. As the table shows, adding levels increases the complexity by $\log_2(k)$, which is better than $k$ of the previous circuit. Note that the number of stages cannot be modified after circuit fabrication.

After the analog divider, there are a few steps before calculating the final quotient voltage. First, the bits need to be converted into a voltage representing the number of levels. To do this, the circuit uses comparators to act as switches to convert each bit to a voltage of either $V_R$ or ground. Then, a summing amplifier (with gains corresponding to the respective powers of 2) calculates the total voltage. Now, we need to add this voltage to one of the VCVS voltages. Both Type-1 and Type-2 VCVS have a comparator at their outputs. If the last bit of the analog divider is 0 (i.e., quotient is even), then Type-1 VCVS voltage is added and vice-versa. 
Although the number of stages cannot be adjusted, $\Delta_H$ (and, thus, the number of levels) can be altered through changing $V_{REF}$. $V_{REF}$ has a minimum of $3~\rm{V}$ since the input $V_{H0}$ reaches up to $3~\rm{V}$ and a maximum of $4.5~\rm{V}$ due to saturation at the Op Amp. This allows increasing $\Delta_H$ up to 50\% from the minimum, which is a decrease in levels up to 33\%. If the number of levels is decreased, $V_R$ can be increased if desired to keep $V_R\times {2^k}$ approximately $V_{DD}$, although it is not required. To make $V_R$ adjustable, however, we need to calculate $V_R=\frac{V_{DD}}{2^k}$, which will require three more Op Amps. To save on power consumption, the proposed circuit keeps $V_R$ constant.

\textbf{LTSpice Simulation:}
Fig.~\ref{fig:LevelsPlot} shows LTSpice simulation output of the circuit's output voltage (AJSCC encoded voltage) for all possible values of $V_{H}$ while fixing $V_{T}$ at 2.5 V so that the mapped point will always be at the center of each level. This plot is generated using $LTC1047$ for the Op Amps and $LTC1441$ for the Comparators instead of $LM324$ and $LP2901$, the respective components used in Design~1~\cite{Zhao16}. These parts are chosen to ensure that the output voltage does not saturate before reaching $V_{DD}$. The figure shows all the 16 levels possible with $k=4$ and matches the expectations of the circuit. The simulation for one stage of ADB is shown in Fig.~\ref{fig:ModulusPlot}. Here, $V_{in}$ and $V_{out}$ are shown to be consistent with Fig.~\ref{fig:ADB2}. Fig.~\ref{fig:ModulusPlot} also shows $V_{comp}$, the bit output from the stage's comparator. $V_{comp}$ represents the modulo quotient of the stage. Note that $V_{comp}$ is high when $V_{in}$ is negative and vice-versa. $V_{out}$ and $V_{comp}$ are both very linear which is important to ensure that circuit calculates the correct level. All other parts of the circuit are (nearly) linear with respect to the input voltages, $V_{H0}$, $V_{T0}$, largely due to the use of Op Amps.

\textbf{Power Consumption:}
Fig.~\ref{fig:PowerPlot} shows the simulated power consumption for different numbers of levels of the ADB circuit with $V_{REF}=3~\rm{V}$ and $V_{DD}=5~\rm{V}$. The circuit here used the same Op Amp $LM324$ and Comparator $LP2901$ as in the parallel VCVS circuit for comparison purposes. At $k=4$ stages/16 levels, the ADB circuit requires $21.66~\mathrm{mW}$, whereas the parallel VCVS circuit requires $22.72~\mathrm{mW}$, which is nearly the same power. For fewer levels, the parallel VCVS circuit is more efficient in power used per level. For the case of designs with $\geq 16$ levels, the ADB-based design is much more (exponentially) power efficient than the parallel VCVS design (see Fig.~\ref{fig:PowerPlot}), e.g., for 128 levels, the power consumption per level of the ADB design is $0.252~\mathrm{mW}$, whereas that of parallel VCVS design is $1.215~\mathrm{mW}$ which is $5\times$ more. 
\begin{table}[t]\scriptsize
 \caption{Power consumption in each module of circuit with non-power efficient components, $k=4$; all values are in $\rm{mW}$.}\label{tab:power}
\begin{center}
 \begin{tabular}{|c| c| c|}
 \hline
\textbf{Subcircuit} & \textbf{Power [mW]} & \textbf{\% of Total}\\ [0.5ex]
 \hline\hline
Analog divider & 14.710 & 67.93\%\\
 \hline
$V_T$ Offset & 0.855 & 3.95\%\\
\hline
 VCVS Type 1,2 & 0.963 & 4.45\%\\
 \hline
 VCVS Type Selector & 0.002 &  0.01\%\\
 \hline
 Bits to Voltage Converter & 3.396 &  15.68\%\\
 \hline
 Final Adder & 0.834 &  3.85\%\\
 \hline
 $V_R$ Buffer & 0.897 &  4.14\%\\
 \hline\hline
 \textbf{Total Circuit} & \textbf{21.660} &  \textbf{100}\textbf{\%} \\
 \hline
\end{tabular}
\end{center}
\end{table}
Table~\ref{tab:power} shows the percentages of power consumption of a 16-level ADB circuit broken down by subcircuits as labeled in Fig.~\ref{fig:prop_ckt_2}. The ADB subcircuit is the most power consuming block, while the bits to voltage converter is the second most. The analog divider is the only subcircuit to significantly increase in power as more stages are added, while the power consumption of the remaining subcircuits remains virtually constant. Hence, in future designs, the ADB subcircuit needs to be made more power efficient. It is worth noting that more power-efficient COTS components can be used to further lower the power consumption. Figure~\ref{fig:PowerPlot} also shows (as ``Efficient ADB'') the power consumption after swapping the Op Amps and Comparators to more power-efficient ones. Here, $LTC1047$ is used for the Op Amp and $LTC1441$ is used for the Comparator. With these efficient components, the ADB circuit consumes $4.8~\rm{mW}$ for 16 levels.

\begin{table}[t]\scriptsize
\caption{Quantity of components used in analog divider vs. parallel VCVS designs, where O=Op Amps, C=Comparators, M=Multiplexers, R=Resistors.}\label{tab:comp}
\begin{center}
 \begin{tabular}{|c|c|c|c|c|c|c|c|c|}
 \hline
 \multicolumn{2}{|c|}{} & \multicolumn{3}{|c|}{\textbf{Analog Divider}} & \multicolumn{4}{|c|}{\textbf{Parallel VCVS}} \\
 \hline
 \textbf{k} & \textbf{Max Levels} & \textbf{\#O} & \textbf{\#C} & \textbf{\#R} & \textbf{\#O} & \textbf{\#C} & \textbf{\#M} & \textbf{\#R}\\ [0.5ex]
 \hline\hline
  1 & 2 & 6 & 4 & 30 & 6 & 2 & 2 & 28\\
 \hline
 2 & 4 & 7 & 6 & 37 & 10 & 4 & 4 & 47\\
 \hline
 3 & 8 & 8 & 8 & 44 & 16 & 8 & 8 & 79\\
 \hline
 4 & 16 & 9 & 10 & 51 & 30 & 16 & 16 & 149\\
 \hline
 5 & 32 & 10 & 12 & 58 & 56 & 32 & 32 & 283\\
 \hline
 6 & 64 &  11 & 14 & 65 & 110 & 64 & 64 & 559\\
 \hline
 7 & 128 & 12 & 16 & 72 & 217 & 128 & 128 & 1103\\
 \hline
\end{tabular}
\end{center}
\end{table}

\textbf{Cost/Component/Complexity Comparison:}
In order to compare the cost of fabrication/complexity of circuit, the number of components are compared in both designs, as shown in Table~\ref{tab:comp}. For the ADB-based design, the number of Op Amps, high power Comparators, low power Comparators, and the Resistors are counted. For the parallel-VCVS design, the number of Op Amps, Comparators, Multiplexers, and Resistors are counted.
From Table~\ref{tab:comp}, we can observe that, with the increasing number of stages, the number of components grows linearly for the ADB-based design with respect to $k$, the logarithmic value of number of stages. In contrast, the number of components grows exponentially for the parallel-VCVS design with respect to $k$. This is the case for all of the components. For example, by comparing 32 vs. 64 vs. 128 levels, it can be observed that, for the ADB-based design, the number of Comparators and Resistors increases linearly for each step; however, for parallel VCVS-based design, the number of Comparators, Multiplexers, and Resistors doubles for each step. This observation indicates that the analog-divider-based design has much lower circuit complexity and hence cost than parallel-VCVS based design.

\textbf{Nano-meter Design Power Consumption:}
An estimate of the power consumption of our circuit when it is fabricated using the latest nano-meter Silicon technology can be made as follows (for the sake of illustration, we will consider 64 levels; a similar analysis can be done for other number of levels too). Based on Table~\ref{tab:comp}, our ADB-based circuit in total has 11 Op Amps, 14 Comparators and 65 Resistors, where Op Amps are clearly the major contributors to the overall power consumption. Using the same nano-designs we considered for these components in~\cite{Zhao16} ($8~\mu\rm{W}$ for Op Amp and $12.7~\rm{nW}$ for comparator), the power consumption of the circuit ($k=6$) can be estimated as $90~\mu \rm{W}$. In contrast, for Design~1 with 11 levels, the estimated power consumption is $130~\rm{\mu W}$.

\balance

\section{Conclusions}\label{sec:conc}
A low-power and low-complexity/cost all-analog circuit is proposed for the realization of rectangular Analog Joint Source Channel Coding~(AJSCC) mapping. The design, which allows tunable spacing between AJSCC levels, is based on Analog Divider Blocks~(ADB) 
and enables inexpensive realizations of wireless sensor nodes for real-time, high temporal/spatial resolution persistent monitoring of urban infrastructure, transportation systems, and precision agriculture, just to name a few applications. Compared with our previous design using parallel Voltage Controlled Voltage Source~(VCVS), this new design based on ADB has a much lower circuit complexity.

\bibliographystyle{IEEEtran}
\bibliography{ref_ajscc_sensor}

\begin{thebibliography}{10}
\providecommand{\url}[1]{#1}
\csname url@rmstyle\endcsname
\providecommand{\newblock}{\relax}
\providecommand{\bibinfo}[2]{#2}
\providecommand\BIBentrySTDinterwordspacing{\spaceskip=0pt\relax}
\providecommand\BIBentryALTinterwordstretchfactor{4}
\providecommand\BIBentryALTinterwordspacing{\spaceskip=\fontdimen2\font plus
\BIBentryALTinterwordstretchfactor\fontdimen3\font minus
  \fontdimen4\font\relax}
\providecommand\BIBforeignlanguage[2]{{%
\expandafter\ifx\csname l@#1\endcsname\relax
\typeout{** WARNING: IEEEtran.bst: No hyphenation pattern has been}%
\typeout{** loaded for the language `#1'. Using the pattern for}%
\typeout{** the default language instead.}%
\else
\language=\csname l@#1\endcsname
\fi
#2}}

\bibitem{Zhao16}
X.~Zhao, V.~Sadhu, and D.~Pompili, ``Low-power all-analog circuit for
  rectangular-type analog joint source channel coding,'' in \emph{IEEE
  International Symposium on Circuits and Systems (ISCAS)}, Montreal, Canada,
  May 2016.

\bibitem{wons3tier2017}
V.~Sadhu, X.~Zhao, and D.~Pompili, ``Energy-efficient analog sensing for
  large-scale, high-density persistent wireless monitoring,'' in \emph{13th
  Annual Conference on Wireless On-demand Network Systems and Services (WONS)},
  Feb 2017, pp. 61--68.

\bibitem{Kumar16}
S.~Kumar, A.~Deshpande, S.~S. Ho, J.~S. Ku, and S.~E. Sarma, ``Urban street
  lighting infrastructure monitoring using a mobile sensor platform,''
  \emph{IEEE Sensors Journal}, vol.~16, no.~12, pp. 4981--4994, June 2016.

\bibitem{Hu15}
X.~Hu, L.~Yang, and W.~Xiong, ``A novel wireless sensor network frame for urban
  transportation,'' \emph{IEEE Internet of Things Journal}, vol.~2, no.~6, pp.
  586--595, Dec 2015.

\bibitem{iscas17}
X.~Zhao, V.~Sadhu, T.~Le, D.~Pompili, and M.~Javanmard, ``Towards low-power
  wearable wireless sensors for molecular biomarker and physiological signal
  monitoring,'' in \emph{IEEE International Symposium on Circuits and Systems
  (ISCAS)}, May 2017, pp. 1--4.

\bibitem{Viswanathan11}
H.~Viswanathan, E.~K. Lee, and D.~Pompili, ``Self-organizing sensing
  infrastructure for autonomic management of green datacenters,'' \emph{IEEE
  Network}, vol.~25, no.~4, pp. 34--40, July 2011.

\bibitem{joe17}
X.~Zhao, D.~Pompili, and J.~Alves, ``Underwater acoustic carrier aggregation:
  Achievable rate and energy-efficiency evaluation,'' \emph{IEEE Journal of
  Oceanic Engineering}, vol.~PP, no.~99, pp. 1--14, 2017.

\bibitem{Magno16wakeup}
M.~Magno, V.~Jelicic, B.~Srbinovski, V.~Bilas, E.~Popovici, and L.~Benini,
  ``Design, implementation, and performance evaluation of a flexible
  low-latency nanowatt wake-up radio receiver,'' \emph{IEEE Transactions on
  Industrial Informatics}, vol.~12, no.~2, pp. 633--644, April 2016.

\bibitem{Basu17}
G.~M. D.~Basu, G. S.~Gupta and X.~Gui, ``Novel adaptive transmission protocol
  for mobile sensors that improves energy efficiency and removes the limitation
  of state based adaptive power control protocol ({SAPC}),'' \emph{Journal of
  Sensor and Actuator Networks}, vol.~6, no.~1, 2017.

\bibitem{Olsson16}
R.~H. Olsson, R.~B. Bogoslovov, and C.~Gordon, ``Event driven persistent
  sensing: Overcoming the energy and lifetime limitations in unattended
  wireless sensors,'' in \emph{IEEE SENSORS}, Oct 2016, pp. 1--3.

\bibitem{Magno16gas}
M.~Magno, V.~Jelicic, K.~Chikkadi, C.~Roman, C.~Hierold, V.~Bilas, and
  L.~Benini, ``Low-power gas sensing using single walled carbon nano tubes in
  wearable devices,'' \emph{IEEE Sensors Journal}, vol.~16, no.~23, pp.
  8329--8337, Dec 2016.

\bibitem{Ahmed16}
K.~Z. Ahmed and S.~Mukhopadhyay, ``A 190 {nA} bias current 10 {mV} input
  multistage boost regulator with intermediate-node control to supply {RF}
  blocks in self-powered wireless sensors,'' \emph{IEEE Transactions on Power
  Electronics}, vol.~31, no.~2, pp. 1322--1333, Feb 2016.

\bibitem{Shannon49}
C.~Shannon, ``Communication in the presence of noise,'' \emph{Proceedings of
  the IRE}, 1949.

\bibitem{mass17}
X.~Zhao, V.~Sadhu, and D.~Pompili, ``Analog signal compression and multiplexing
  techniques for healthcare internet of things,'' in \emph{14th International
  Conference on Mobile Ad Hoc and Sensor Systems (MASS)}, Oct 2017.

\bibitem{Hekland05}
F.~Hekland, G.~Oien, and T.~Ramstad, ``Using 2:1 {S}hannon mapping for joint
  source-channel coding,'' in \emph{Data Compression Conference (DCC)}, March
  2005, pp. 223--232.

\bibitem{Fresnedo15}
O.~Fresnedo, J.~Gonzalez-Coma, M.~Hassanin, L.~Castedo, and J.~Garcia-Frias,
  ``Evaluation of analog joint source-channel coding systems for multiple
  access channels,'' \emph{IEEE Transactions on Communications}, vol.~63,
  no.~6, pp. 2312--2324, June 2015.

\bibitem{Brante13}
G.~Brante, R.~Souza, and J.~Garcia-Frias, ``Spatial diversity using analog
  joint source channel coding in wireless channels,'' \emph{IEEE Transactions
  on Communications}, vol.~61, no.~1, pp. 301--311, January 2013.

\bibitem{Garcia11}
J.~Garcia-Naya, O.~Fresnedo, F.~Vazquez-Araujo, M.~Gonzalez-Lopez, L.~Castedo,
  and J.~Garcia-Frias, ``Experimental evaluation of analog joint source-channel
  coding in indoor environments,'' in \emph{IEEE International Conference on
  Communications (ICC)}, June 2011, pp. 1--5.

\bibitem{Romero14}
S.~Romero, M.~Hassanin, J.~Garcia-Frias, and G.~Arce, ``Analog joint source
  channel coding for wireless optical communications and image transmission,''
  \emph{Journal of Lightwave Technology}, vol.~32, no.~9, pp. 1654--1662, May
  2014.

\bibitem{stopler14}
D.~Stopler, ``Device method and system for communicating data,'' Jan 2014, {US}
  Patent 8,625,709.

\bibitem{foldingadc}
{Analog Devices}, ``{MT-025 Tutorial},''
  \url{http://www.analog.com/media/en/training-seminars/tutorials/MT-025.pdf}.

\end{thebibliography}

\end{document}